\documentclass[9pt,journal]{IEEEtran}
\usepackage{authblk}
\usepackage[utf8]{inputenc}
\usepackage{caption}
\usepackage{graphicx}
\usepackage{subcaption}
\usepackage[dvipsnames]{xcolor}
\usepackage{epstopdf}

\title{Reconfigurable Intelligent Surfaces: Potentials, Applications, and Challenges for 6G Wireless Networks}

\author{Sarah Basharat, Syed Ali Hassan, Haris Pervaiz, Aamir Mahmood, Zhiguo Ding, and Mikael Gidlund

\thanks{S. Basharat and S. A. Hassan  are with the School of Electrical Engineering
and Computer Science (SEECS), National University of Sciences and Technology (NUST), Islamabad 44000, Pakistan.}
\thanks{H. Pervaiz is with the School of Computing and
Communications, Lancaster University, Lancaster LA1 4YW, U.K.}
\thanks{A. Mahmood and M. Gidlund are with Mid Sweden University, Sweden.}
\thanks{Z. Ding is with the School of Electrical and Electronic Engineering, The University of Manchester, Manchester M13 9PL, U.K.}%
}

\usepackage{amsmath}
\usepackage{amssymb}
\usepackage{cite}
\usepackage{multirow}
\usepackage{tabularx}
\newcolumntype{P}[1]{>{\centering\arraybackslash}p{#1}}
\newcolumntype{M}[1]{>{\centering\arraybackslash}m{#1}}
\begin{document}
\maketitle
\begin{abstract}
Reconfigurable intelligent surfaces (RISs), with the potential to realize smart radio environment, have emerged as an energy-efficient and a cost-effective technology to support the services and demands foreseen for coming decades. By leveraging a large number of low-cost passive reflecting elements, RISs introduce a phase-shift in the impinging signal to create a favorable propagation channel between the transmitter and the receiver.~\textcolor{black}{In this article, we provide a tutorial overview of RISs for sixth-generation (6G) wireless networks. Specifically, we present a comprehensive discussion on performance gains that can be achieved by integrating RISs with emerging communication technologies. We address the practical implementation of RIS-assisted networks and expose the crucial challenges, including the RIS reconfiguration, deployment and size optimization, and channel estimation. Furthermore, we explore the integration of RIS and non-orthogonal multiple access (NOMA) under imperfect channel state information (CSI). Our numerical results illustrate the importance of better channel estimation in RIS-assisted networks and indicate the various factors that impact the size of RIS. Finally, we present promising future research directions for realizing RIS-assisted networks in 6G communication.}
%which transforms the propagation channel into controllable system block that can be optimized for overall performance enhancement of the communication system.
\end{abstract}
\section{Introduction}
\IEEEPARstart{T}{he} global mobile traffic volume is anticipated to reach 5016 exabytes per month (Eb/mo) in 2030, which was 7.462 EB/mo in 2010~\cite{6G}. This clearly depicts the importance of the evolution and advancement of mobile communication technologies. To date, fifth-generation (5G) communication, which is expected to realize the targeted 1000x increase in network capacity with new and advanced services, is being rolled out in the world. However, 5G systems will not be able to fully support the growing demand for wireless communication in 2030. The core 5G technologies  include massive multiple-input multiple-output (mMIMO) and millimeter-wave (mmWave) communications. The mMIMO technology exploits the spatial domain by deploying numerous antennas to enable parallel transmission to multiple users using the same frequency-time resource block. The mmWaves, on the other hand, offer plenteous spare spectrum in high frequency bands, thus resolving spectrum scarcity issues at microwave frequencies. Although mMIMO and mmWaves significantly improve spectral efficiency (SE), high hardware cost and complexity are  major hurdles in their practical implementation. Moreover, mmWaves are highly vulnerable to signal blockage and attenuation. Therefore, reliable and efficient communication is still not guaranteed.

While 5G is yet to be realized fully, the researchers have already started looking for energy and spectral-efficient solution for sixth-generation  (6G) systems. In addition to the energy and spectral efficiency, the new paradigm is smart and reconfigurable wireless environments~\cite{RuiZhang}. Recently, a cost-effective and energy-efficient technology, reconfigurable intelligent surface (RIS)~\cite{RIS_UAV1, IRS_mmWave,IRS_MIMO}, also called  intelligent reflecting surface (IRS)~\cite{RuiZhang,bjornson2020power,IRSNOMA_ZDing,IRSNOMA_opt,RIS_SWIPT,RIS_UAV2,IRS_backCom1,IRS_backCom2,RIS_NOMA_mmWave,Channel_etimation2}, or passive holographic MIMO surface (HMIMOS)~\cite{passive_HMIMOS}, has been proposed in this regard. RIS empowers the smart wireless environments by overcoming the stochastic nature of the propagation channel, thereby improving quality-of-service (QoS) and connectivity. The wireless environment that used to be a dynamic uncontrollable factor is now considered to be a part of the network design parameter.

Specifically, an RIS is a software-controlled planer surface consisting of a large number of low-cost passive reflecting elements. Each element, of size smaller than the wavelength, has the capability to alter the phase of the impinging signal, creating a favorable wireless environment between the transmitter and the receiver. \textcolor{black}{The RIS reflection adaption is programmed and controlled via a smart controller, such as a field-programmable gate array (FPGA), which acts as a gate-way to communicate and coordinate with the BS through a separate wireless or a wired link.} In particular, RIS receives a signal from the base station (BS), and then reflects the incident signal by inducing phase changes, adjusted by the controller. Consequently, the reflected signal can be added coherently with the direct signal from the BS to either boost or attenuate the overall signal strength at the receiver. \textcolor{black}{Although RIS is theoretically passive, since it reflects the signal without power amplification, however, in practice, it has minimal power requirement for the operation of smart controller and reconfiguration of the elements for controllable reflections.}

\begin{figure*}[t]
\centering   
\includegraphics[width=0.98\textwidth]{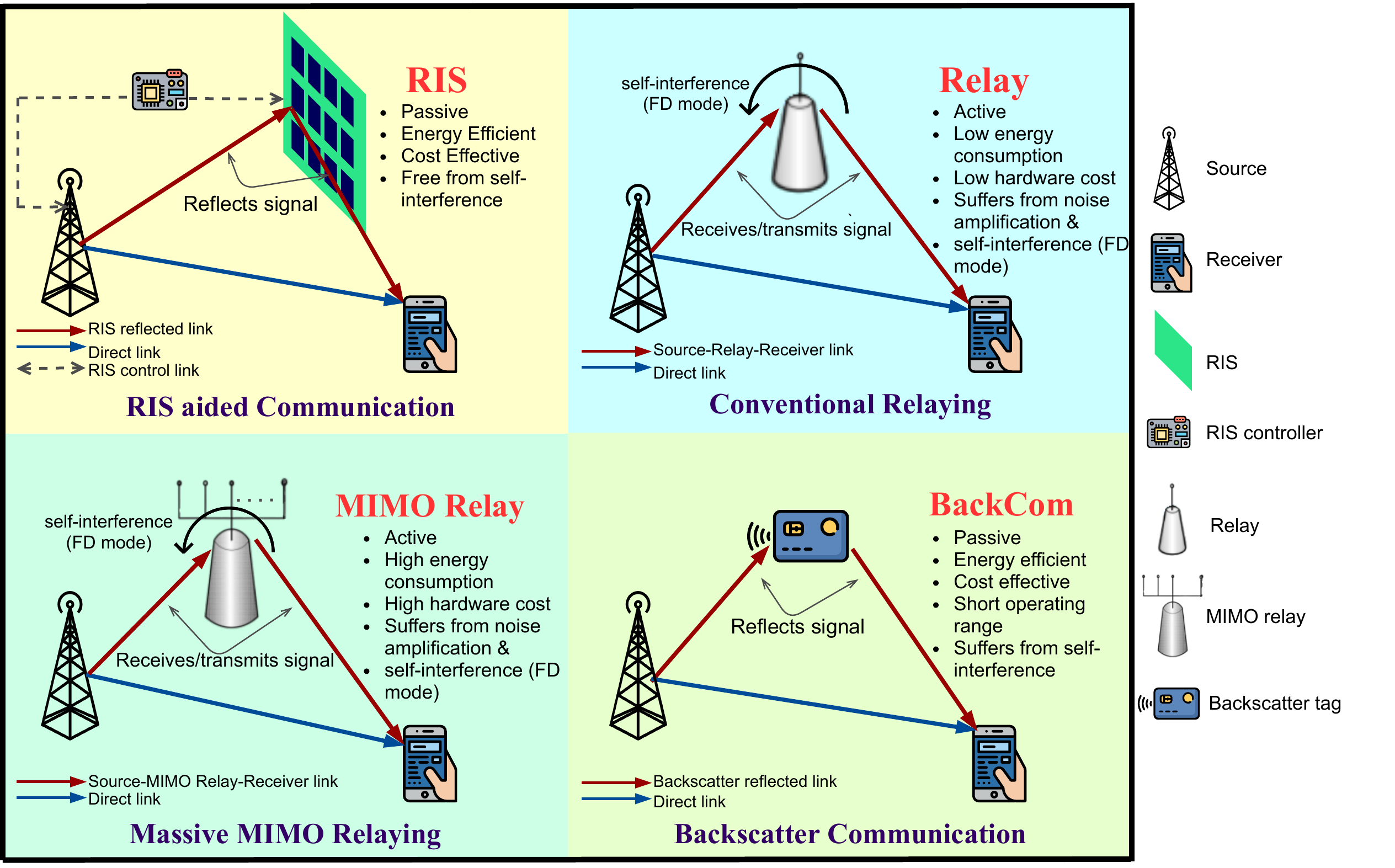}
\caption{RIS comparison with existing related technologies.}\label{fig.1}
\end{figure*}

\textcolor{black}{As illustrated in Fig.~\ref{fig.1}, RIS concept can be viewed to operate similarly as other related wireless technologies such as, conventional relaying, backscatter communication (BackCom), and mMIMO relaying. We now present the major differences and competitive strengths of RIS that make it stand out among these technologies. First, compared to conventional relaying that requires additional power for signal transmission, amplification, and regeneration, RIS passively reflects the impinging signal by inducing intelligent phase-shifts, without the need for an additional radio frequency (RF) source.  Moreover, RIS operates in full-duplex (FD) mode, free from noise amplification and self-interference. Secondly, compared to traditional BackCom such as radio frequency identification (RFID) tag that loads its own information on the incident signal and then backscatters the modulated signal to the receiver, RIS reflects the incident signal to assist the communication between the source and the receiver without sending any information of its own. BackCom also requires sophisticated signal processing for self-interference cancellation in order to decode the tag message. Finally, unlike mMIMO relaying, RIS can be implemented at a much low hardware cost and power consumption. Although signal-to-noise ratio (SNR) achieved through RIS is less than the equal-sized mMIMO counterpart~\cite{bjornson2020power}, however, the SNR of RIS-assisted system can be improved by increasing the reflecting elements, since the cost per reflecting element of RIS is much less than the cost per antenna element in mMIMO relaying. Architecturally, RIS is lightweight with conformal geometry and can be easily mounted on the ceilings, walls, and building facades.}

Inspired by the RIS potential to realize smart wireless environments and its compatibility with other technologies, the main contributions of this study can be summarized as follows. 
\textcolor{black}{\begin{itemize}
    \item We provide a comprehensive discussion on integrating RIS with emerging communication technologies for realizing 6G wireless networks, namely, non-orthogonal multiple access (NOMA), simultaneous wireless information and power transfer (SWIPT), unmanned aerial vehicles (UAVs), BackCom, mmWaves, and multi-antenna systems.
    \item For practical implementation of RIS-assisted networks, we identify three crucial challenges, including RIS reconfiguration for controllable reflections, deployment and size optimization, and channel estimation.
    \item We present a novel case study for RIS-assisted NOMA network with imperfect channel state information (CSI) to highlight the impact of channel estimation errors on the performance of RIS-assisted NOMA networks. We further determine the various factors that affect the size of RIS, i.e., the number of RIS elements.
    \item To provide effective guidance for future research, we introduce five promising research directions for realizing RIS-assisted networks.
\end{itemize}}

\section{Integrating RIS with emerging communication technologies}
Current research contributions have revealed RIS to be a cutting-edge technology, opening new promising research opportunities on the road towards 6G. 
In this section, we elaborate on the performance gains that can be achieved by integrating RIS with emerging communication technologies.
\textcolor{black}{\subsection{RIS and NOMA}}
 NOMA has emerged as a promising technology for future generation networks to support massive connectivity. Power domain NOMA (PD-NOMA) enables multiple users to share the same resource (e.g., time, frequency, code) block, hence improves both spectral and energy efficiency. In a downlink NOMA system, superposition coding (SC) is used at the BS to multiplex the data of the users with different channel gains, and successive interference cancellation (SIC) is employed at the receiver to decode the message signals. Although NOMA provides sufficient performance gains over OMA, the stringent demands on data rate and connectivity for B5G/6G systems compel to shift to smart and reconfigurable wireless networks; hence, the RIS-assisted NOMA system has been proposed by the research community. 
\vspace{-0.1cm}

To fully exploit the gains of the RIS-assisted NOMA system, appropriate phase-shift matrix and beamforming vectors need to be designed. The two types of phase-shifting designs, \emph{coherent phase-shifting} and \emph{random phase-shifting} achieve different trade-offs between performance and complexity for an RIS-assisted NOMA system~\cite{IRSNOMA_ZDing}. For coherent phase-shifting, each RIS element introduces a phase that matches the phase of fading channels from RIS-to-BS and users. Despite the superior performance, it might not be feasible to implement coherent shifting design because of hardware limitations of phase shifters and excess system overhead of acquiring CSI. Although the random phase-shifting design reduces the system complexity and overhead of acquiring CSI, it degrades the system performance. Furthermore, the RIS phase-shift matrix and beamforming vectors can be jointly optimized to minimize the transmit power of the BS using semi-definite relaxation (SDR). \textcolor{black}{However, for large-scale RIS-assisted networks, the computational complexity of the SDR technique is extremely high and the probability of returning rank-one optimum solution is extremely small.} Therefore, a difference-of-convex (DC) method has been proposed that overcomes the limitation of the SDR method and outperforms in terms of minimizing the BS transmit power~\cite{IRSNOMA_opt}.

\subsection{RIS and SWIPT}
SWIPT is an effective solution to power massive devices in a wireless-powered Internet-of-things (IoT) network. In practice, the significant power loss over long distances reduces the energy harvested at the energy receiver, which degrades the performance of SWIPT systems. However, the limitations of practical SWIPT systems can be compensated via RIS, as illustrated in Fig.~\ref{fig.2}. Through intelligent signal reflections, RIS boosts the signal strength both at the information receiver (IR) and the energy receiver (ER), thereby improving the energy efficiency of SWIPT system.

 Recently,~\cite{RIS_SWIPT} unleashes the benefits of RIS-assisted SWIPT network, where a multi-antenna BS serves several multi-antenna information users, while satisfying the energy requirements of the energy users. The authors proposed a weighted sum-rate maximization problem for the joint optimization of BS's transmit precoding matrices and RIS phase-shifts. The proposed problem is challenging to solve for optimal solution, owing to highly coupled optimization parameters. Therefore, the authors developed an iterative solution using a block coordinate descent (BCD) algorithm that converges rapidly and outperforms the benchmark approaches, i.e., fixed RIS phase-shifts and conventional SWIPT networks without RIS. Such a performance is quite appealing for practical applications.

\subsection{RIS and UAVs}
\begin{figure}[t]
\centering   
\includegraphics[width=0.48\textwidth]{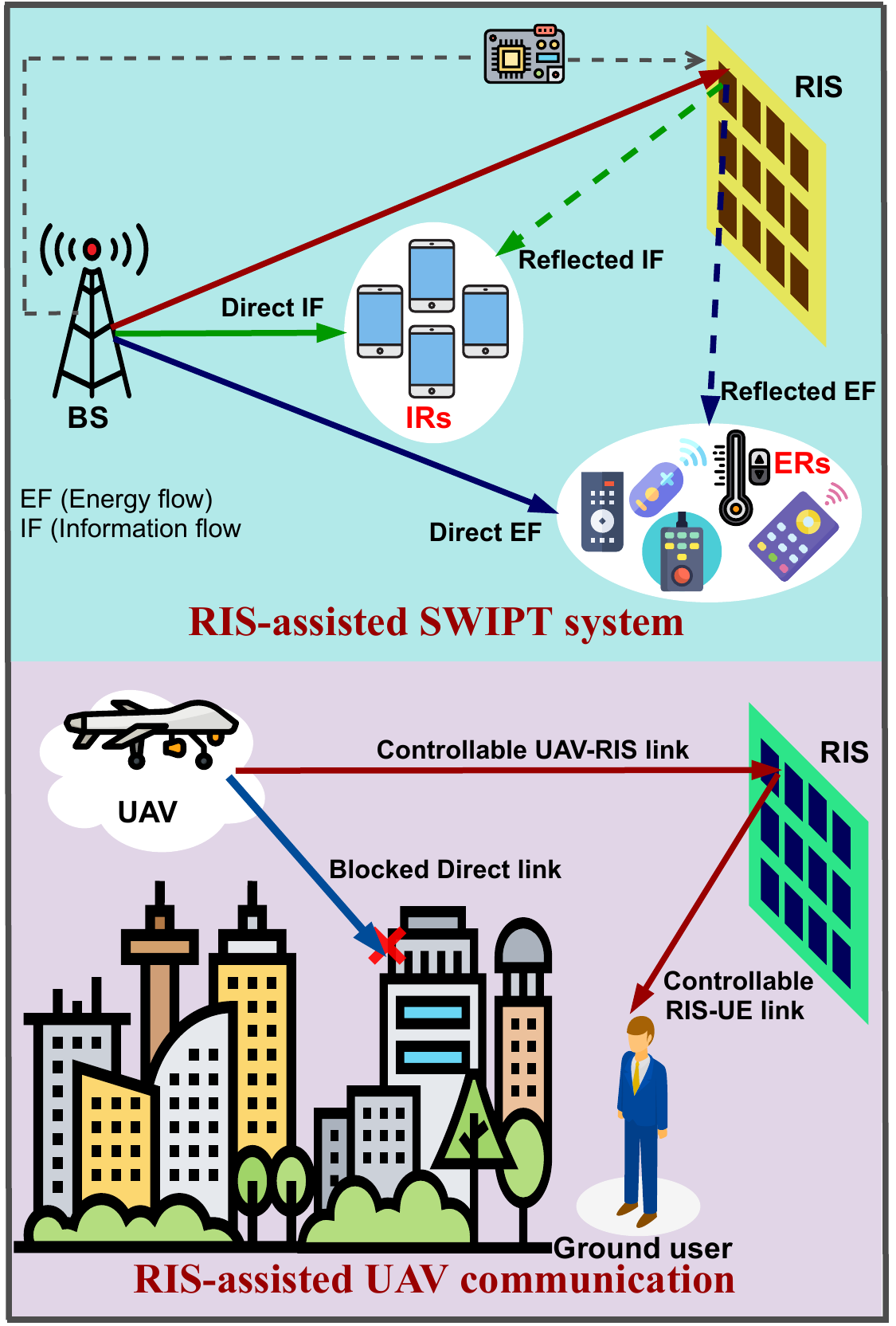}
\caption{RIS-assisted SWIPT and UAV communication.}\label{fig.2}
\end{figure}

RIS can even be applied to UAV-enabled communication systems to improve propagation environment and enhance communication quality, as shown in Fig.~\ref{fig.2}. In a dense urban environment, the line-of-sight (LoS) links between the UAV and the ground users may be blocked, which deteriorates the channel gains. However, the RIS-assisted UAV system can enable virtual LoS paths by reflecting the signal received from UAV to the ground users. The received signal power at the ground user can be significantly enhanced by the joint optimization of RIS beamforming and UAV trajectory. In this regard, Li \emph{et al.}~\cite{RIS_UAV1} proposed a iterative algorithm for achievable rate maximization under UAV mobility and the RIS’s phase-shift constraints, with the results confirming significant improvement in achievable rate with the aid of RIS.

RIS can also enhance the cellular communication of UAVs~\cite{RIS_UAV2}, which suffers from down-tilted BS antennas, i.e., the main lobes of antennas are optimized to serve the ground users, while UAVs communication is supported by side lobes only. Through intelligent and optimized signal reflections, controlled via cellular BS, RIS can direct the impinging BS signal towards a specific UAV. The RIS reflected signal combines coherently with the direct BS-UAV signal, thus improves the received signal strength at the UAV. Even a small-sized RIS, deployed on a building  facade, can improve the cellular communication of the UAV flying substantially above the BS. Moreover, RIS location, i.e., RIS distance from BS distance and RIS deployment height, is a critical factor in such applications as performance gain achieved through RIS is maximized when RIS location is selected optimally. 

\begin{figure*}[t]
\centering   
\includegraphics[width=1\textwidth]{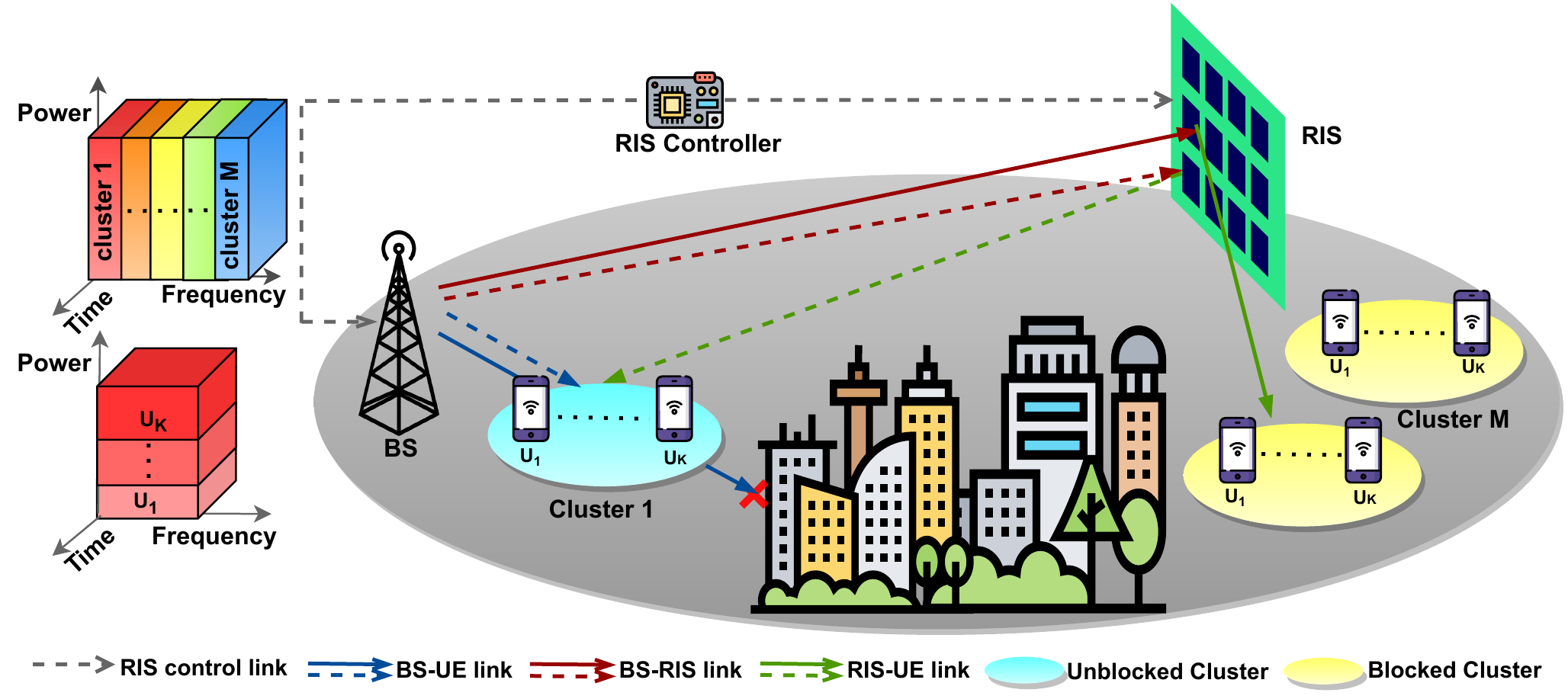}
\caption{An illustration of RIS-assisted downlink NOMA network.}\label{fig.3}
\end{figure*}
\vspace{-0.4cm}
\textcolor{black}{\subsection{RIS and BackCom}}
\textcolor{black}{
BackCom is a promising solution towards an energy-efficient and sustainable IoT network. Despite the extensive research on the improvement of reliability and throughput of a BackCom system, its short operation range remains a key barrier towards the large-scale deployment that needs to be addressed. Recently, in ~\cite{IRS_backCom1}, the authors elaborated on the potentials of RIS-assisted monostatic and bistatic BackCom systems, where the RIS is employed to assist the communication between the tag and the reader. The authors proposed a joint optimization framework, under transmit power minimization, to optimize RIS phase-shifts and the source transmit beamforming. The application of RIS to BackCom can significantly reduce the transmit power, which can be mapped to improve the operational range.}

\textcolor{black}{Furthermore, the ability of RIS to steer the signals in different directions to reduce the inter-user interference can be utilized to improve the detection performance of the ambient BackCom systems. In this regard, Jia \emph{et al.}~\cite{IRS_backCom2} proposed a deep reinforcement learning (DRL) based approach, namely, the deep deterministic policy gradient (DDPG) algorithm, to jointly optimize the RIS and reader beamforming for RIS-assisted ambient BackCom system with no knowledge of channels and ambient signals. The results in \cite{IRS_backCom2} demonstrate the significant improvement in detection performance of ambient BackCom with the aid of RIS.} %DRL based approach outperforms the various full CSI benchmark approaches in terms of detection performance.}

\textcolor{black}{\subsection{RIS and mmWaves}}
\textcolor{black}{The mmWave communication, with the capability to support multi-gigabits of data rate, is perceived as a potential solution for the looming capacity crunch. However, high directivity of mmWaves makes it vulnerable to blockage instants, especially in indoor and dense urban environments. As RIS has the capability to introduce effective additional paths, an RIS-enhanced mmWave system can overcome the limitations of a conventional mmWave system. When the direct links from the BS to users are severely blocked, optimizing the system parameters can provide satisfactory performance gains. Recently, in \cite{RIS_NOMA_mmWave}, the authors utilized the alternative optimization and successive convex approximation (SCA) to jointly optimize the beamforming vectors and power allocation for the RIS-assisted mmWave-NOMA system. The results confirm the RIS's ability to enhance the coverage range of the mmWave-NOMA system, especially when the direct BS to users' links are blocked.}

\textcolor{black}{Furthermore, the hybrid precoding design for multi-user RIS-assisted mmWave communication system is presented in~\cite{IRS_mmWave}, where the direct links from the BS and the users are assumed to be blocked. The authors jointly optimized the hybrid precoding at the BS and phase-shifts at the RIS to minimize the mean square error (MSE) between the transmitted and the received symbols. The gradient-projection (GP) method, based on alternating optimization (AO), is adopted to address the non-convex constraint for the analog precoding and the phase-shifts. The results illustrate the significant performance gains of the proposed design. However, the efficacy of the proposed design under imperfect CSI needs further investigation. Moreover, the proposed design for the hybrid precoding and phase-shifts can be extended to a system with the direct links between the BS and the users. 
}

\textcolor{black}{\subsection{RIS and Multi-antenna Systems}}
\textcolor{black}{The multi-antenna systems aim at actively improving the signal quality by employing a large number of antennas and exploiting the spatial domain for transmit beamforming. However, the conventional multiple-input single-output (MISO) systems suffer from wireless channel randomness, limiting their performance. Therefore, for an energy-efficient solution, RIS can be applied to MISO systems to improve the network performance at significantly low hardware cost and energy consumption. Different from the conventional systems, an RIS-aided MISO system can guarantee the users' QoS, with less number of BS antennas, by utilizing the smart passive reflections of RIS.}

\textcolor{black}{Recently,~\cite{IRS_MIMO} presented a hybrid beamforming design for a multi-user RIS-assisted MISO system, where the communication is established via RIS only due to the presence of obstacles between the BS and the various users. The authors proposed a two-step sum-rate maximization algorithm to design the continuous digital beamforming at the BS and discrete analog beamforming at the RIS. The observations indicated in~\cite{IRS_MIMO} provide useful insights into the design of RIS-based systems. Firstly, the system performance is influenced by the size of RIS and the number of quantization bits for discrete phase-shifts. Secondly, the RIS-based hybrid beamforming design can greatly reduce the requirement of dedicated hardware while providing the satisfactory sum-rate.}
\vspace{0.5cm}
\section{Practical Implementation of RIS-assisted Networks}
In this section, we identify and discuss the crucial challenges for the practical implementation of RIS-assisted networks.

\subsection{\textcolor{black}{RIS Reconfiguration for Controllable Reflections}}
\textcolor{black}{The RIS phase-shift per element can be tuned for controllable reflections through three main approaches, namely, mechanical actuation, functional materials, and electronic devices. Besides the phase control, the reflection amplitude can be adjusted by varying the load impedance in each element. Thus, the reflection amplitude and phase-shift can be realized in the range of \([0, 1]\) and \([0, 2\pi )\), respectively. The continuous variation of reflection coefficients is usually beneficial from the communication performance perspective. However, for practical RISs with a massive number of reflecting elements, it is desirable to implement only a finite number of discrete phase-shift and amplitude levels, since high-resolution elements increase the hardware cost, design complexity, and control overhead. In addition, the optimization of reflection amplitudes and phase-shifts becomes more challenging with discrete variables. Furthermore, most of the works on RIS assume ideal phase-shift model with full signal reflection at each element regardless of the phase-shift at each element, which is difficult to realize in practice because of the strong coupling amongst the reflection amplitude and phase-shift. Therefore, realistic phase-shift models, with phase-dependent amplitude response, have to be conceived for accurate performance analysis. }

\subsection{\textcolor{black}{RIS Deployment and Size Optimization}}
\textcolor{black}{The RIS deployment in a hybrid network, with both active BSs and passive RISs, is another crucial problem. The deployment strategy significantly influences the distribution of the RIS associated propagation channels, i.e., BS-RIS and RIS-users' channels. Therefore, an effective RIS deployment strategy needs to be adopted to guarantee the performance enhancements promised by the RIS technology. Moreover, the RIS deployment must also consider the practical factors, namely, deployment cost, user distribution, and available space, along with propagation conditions. For the single-user design, the given number of RIS elements can be either grouped as a single RIS or partitioned into multiple cooperative RISs for reaping the cooperative passive beamforming gain. Furthermore, for the multi-user design, the given number of RIS elements can be either grouped as a single RIS placed in the vicinity of the BS (centralized deployment) or partitioned into multiple RISs placed closed to users' hot spots (distributed deployment). In general, the distributed RISs have a greater probability of establishing LoS links with the BS and the users than the centralized RIS. However, in distributed deployment, the communication between the RISs and the BS, and the coordination among the multiple RISs greatly increase the signalling overhead. Moreover, how to optimally select the active number of RIS elements in both centralized and distributed placement is another design challenge.  }
\begin{figure}
\centering
\includegraphics[width=0.5\textwidth]{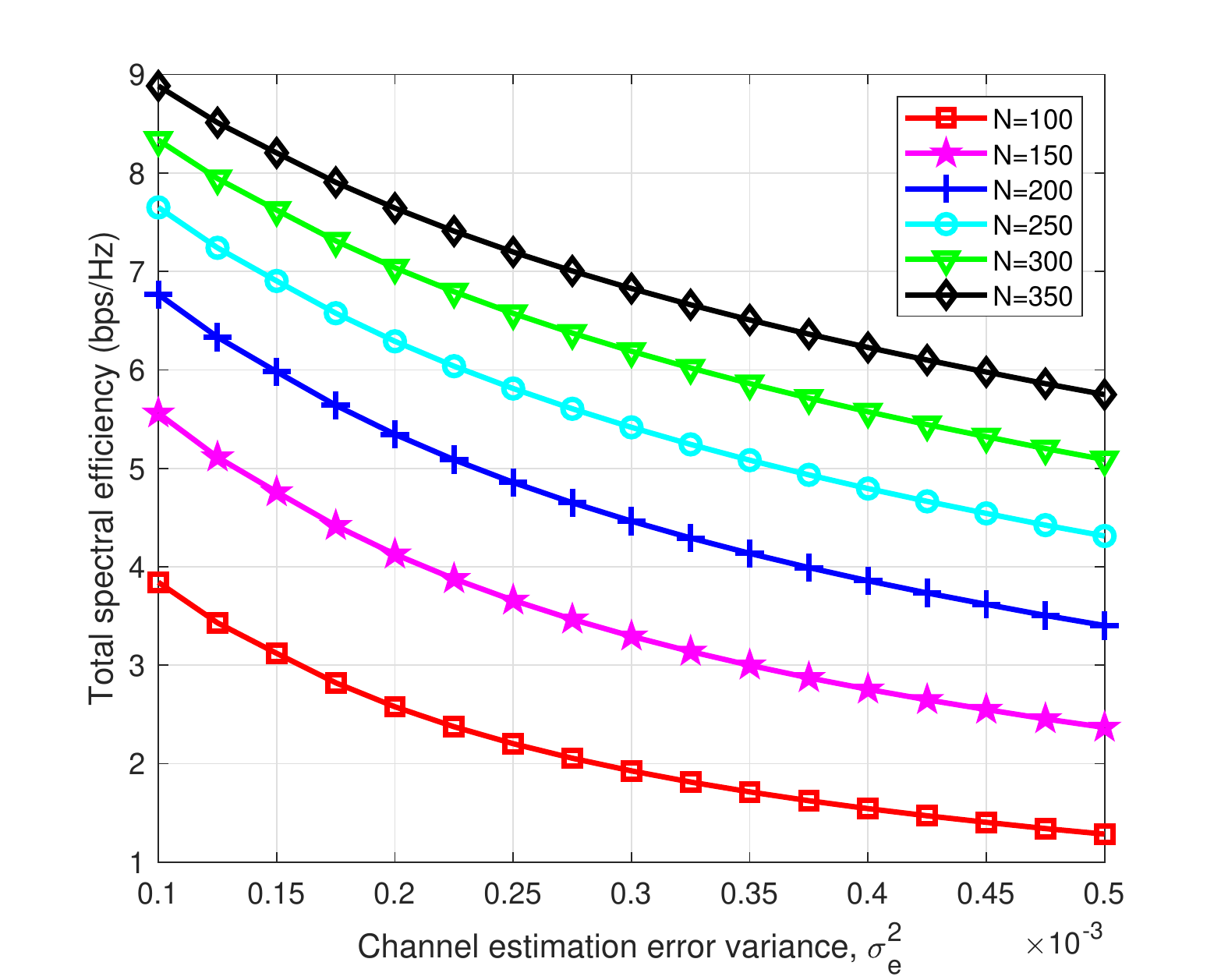}
\caption{The total SE versus channel estimation error variance for different number of RIS elements ($N$) with transmit power ($P_s$) = $30$ dBm, and  users ($K$) = $3$.}\label{fig.4}
\end{figure}

\begin{figure}
\centering   
\includegraphics[width=0.5\textwidth]{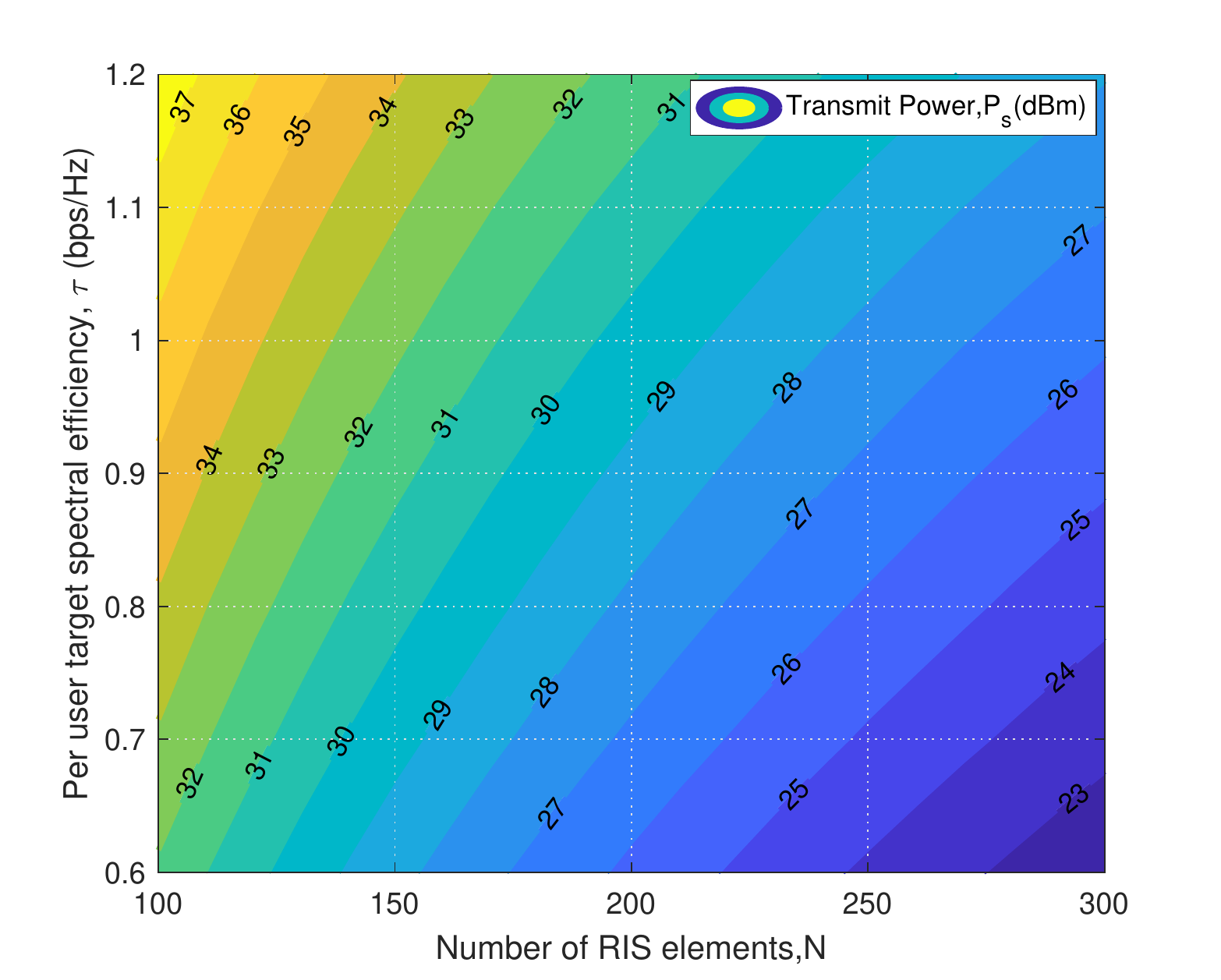}
\caption{Required transmit power for varying per user target SE and number of RIS elements with channel estimation error variance ($\sigma_e^2$) = \(0.0005\), and users ($K$) = \(3\). }
\label{fig5}
\end{figure}

\subsection{Channel Estimation in RIS-assisted Networks}
The channel estimation for RIS-assisted networks is performed at the BS station, and the acquired CSI is communicated to the RIS controller via control link, which adjusts the phase-shifts. One practical approach for RIS’s channel estimation is by employing an element-by-element ON/OFF-based channel estimation scheme, i.e., only one RIS element is turned on each time while all other elements are set OFF, consequently, the direct channels from the BS to users and the RIS reflected channels are estimated separately. However, the ON/OFF-based scheme is not cost-effective for large-scale RISs because of training overhead. Furthermore, the RIS reflected signal suffers from substantial power loss as each time only one element is turned on. This weakens the received signal strength, which degrades the channel estimation accuracy. To reduce the training overhead for practical scale RISs, the RIS elements are grouped into sub-surfaces~\cite{Channel_etimation2}; as a result, only the RIS reflected channel associated with the sub-surface needs to be estimated. However, the design of low-overhead channel estimation protocols for future networks with a large number of users remains an open problem. Moreover, most of the existing  works on RIS assume the availability of perfect CSI at the BS, which is practically not feasible. Therefore, the channel estimation errors should be taken into account for accurate performance analysis. 

%The research contributions on the RIS deployment strategies suffer from paucity, which thus demands for future work.  }

\section{RIS-assisted NOMA Network under imperfect CSI: A case study}
The integration of RIS with NOMA can provide a potential multiple access solution for future networks. RIS-assisted NOMA networks can play a significant role in improving  network coverage and capacity in urban areas where high-rise buildings and structures disrupt wireless services. As illustrated in Fig.~\ref{fig.3}, we consider an RIS-assisted downlink PD-NOMA network, where a macro BS is equipped with a single transmit antenna and an RIS equipped with \(N\) reconfigurable passive reflecting elements serve uniformly distributed single-antenna users, grouped into clusters. The RIS  is connected to a smart controller, which communicates with the BS and
changes the phase of the incident signal based on coherent phase-shifting design~\cite{IRSNOMA_ZDing}. The total system bandwidth is equally divided into orthogonal frequency
resource blocks, such that each NOMA cluster is allocated a single frequency
resource block. The available BS power is equally divided among the resource blocks, such that the power allocated to a single resource block is \(P_s\). In the considered network, the direct links from the BS to users are blocked for some clusters, due to the blocking objects, hence such users exploit RIS to establish communication. On the other hand, some clusters have strong and direct BS-to-user links such that the signal received via RIS is negligible because of the increased path
loss.

To better explore the role of RIS, we focus on a single blocked cluster with \(K\) users, where we assumed the user \(1\) and user \(K\) to be the strongest and the weakest users, respectively, based on their overall channel gains. Following the fixed power allocation approach for NOMA users, the highest channel gain user, i.e., the strongest user, gets the minimum share of transmit power while the lowest channel gain user, i.e., the weakest user, gets the maximum share of the transmit power. All wireless links, i.e., BS to RIS link and RIS to users' links, are modeled as Rayleigh fading channels with path loss and
perturbed by additive white Gaussian noise (AWGN). We modeled the channel estimates using the minimum  mean  square  error (MMSE) channel estimation error model, whose quality of estimation is indicated by the variance of the channel estimation error, \(\sigma^2_e\), smaller the error variance better the estimation. The channel estimation error is regarded as interference in the system, which adversely affects the system performance. 
%Hence, the performance of RIS-assisted networks highly depends on the quality of channel estimation. Better the estimation, higher the performance gains.

\subsection{Impact of Imperfect CSI and RIS Elements}
The SE, i.e., the ratio of achievable rate and bandwidth, in the RIS-assisted NOMA network highly depends on the quality of channel estimation and the number of RIS elements. In Fig.~\ref{fig.4} we demonstrate the impact of varying channel estimation error variance, and the number of RIS elements on total SE, i.e., the SE achieved by \(K\) users. For any value of RIS elements, as error variance increases, the total SE decreases, since the channel estimation error acts as a source of interference. However, even with imperfect CSI, high SE can be achieved by deploying a higher number of RIS elements. 
%This is due to fact that increasing number of RIS elements add beamforming gain which increases the signal-to-interference plus-noise ratio (SINR), thus improving SE.

The transmit power consumption for RIS-assisted NOMA network depends upon the number of RIS elements, and the target spectral efficiency. In this regard, Fig.~\ref{fig5} shows the impact of varying per user target spectral efficiency and the number of RIS elements on BS transmit power. First, it is observed that for a given number of RIS elements, the required transmit power increases with the increase in target spectral efficiency. Second, the transmit power scales down with the increase in the number of RIS elements. For instance, for the same per-user target spectral efficiency of $1$ bps/Hz, a transmit power of $32$ dBm is required for $150$ RIS elements, while this value reduces to about $26$ dBm for $300$ elements, which indicates $6$ dB gain by doubling the RIS elements. From this, we can conclude that RIS passive reflection adds power gain, which can be either utilized to improve SE or reduce total power consumption.

\subsection{Factors Affecting the Size of RIS}
\begin{figure}[t]
\centering   
\includegraphics[width=0.5\textwidth]{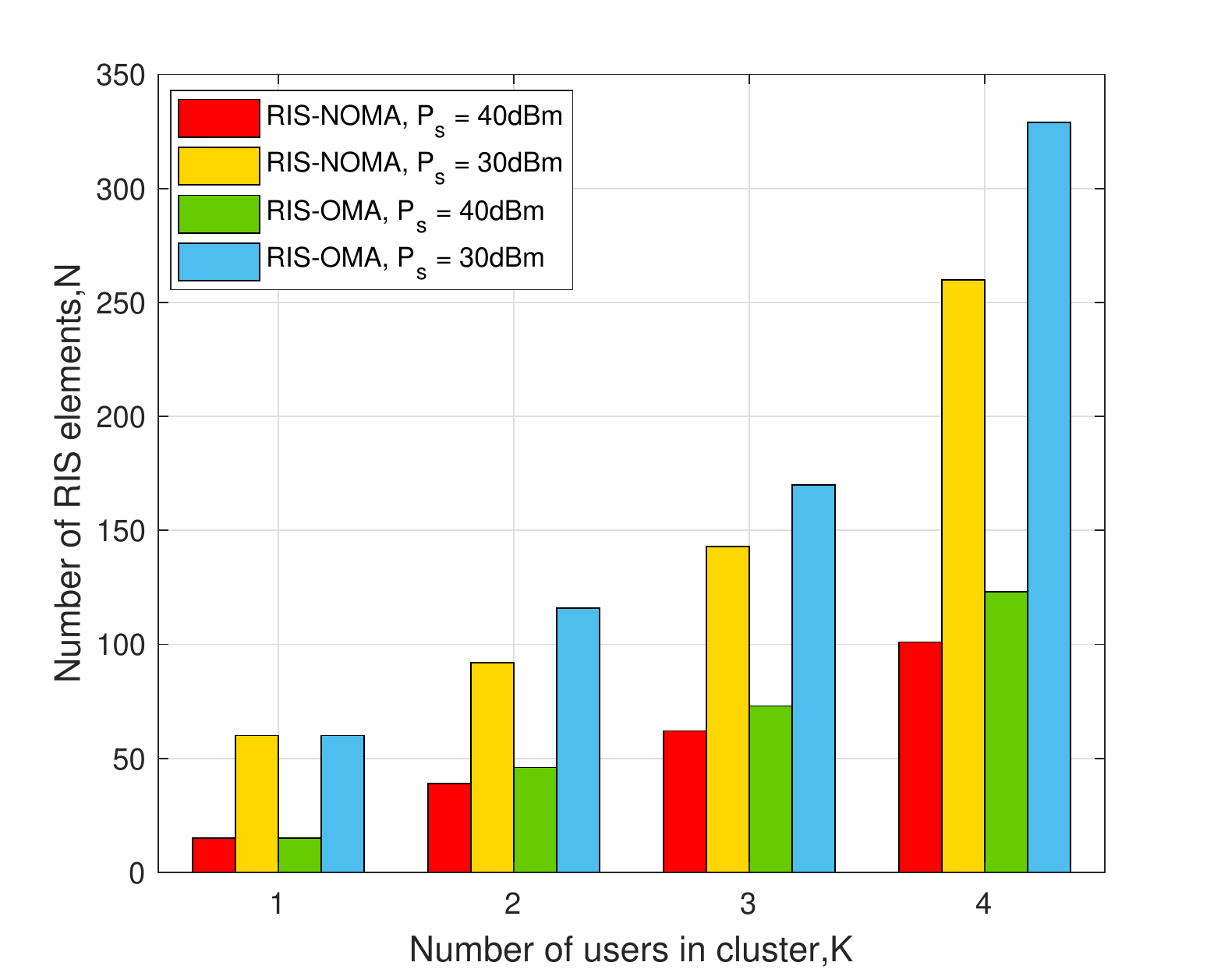}
\caption{\textcolor{black}{Number of RIS elements versus the number of users for RIS-NOMA and RIS-OMA with per user target SE ($\tau$) = \(1.2\) bps/Hz, and channel estimation error variance ($\sigma_e^2$) = \(0.0001\).}}\label{fig.6}
\end{figure}

\textcolor{black}{The minimum number of RIS elements required to guarantee the per user target SE varies with many factors. In Fig.~\ref{fig.6}, we compare the number of RIS elements required for RIS-NOMA and RIS-OMA with varying number of users and BS transmit power. Here, for RIS-OMA, we consider frequency division multiple access (FDMA), where multiple users transmit data simultaneously at different frequency slots. It can be observed that for the same number of users and transmit power, the number of RIS elements required to achieve the target SE are greater for RIS-OMA than for RIS-NOMA. This clearly depicts that the superiority of NOMA compared to OMA still remains after introducing the RIS. The number of RIS elements also increases with the increase in number of users in the cluster, owing to the increase in required orthogonal sub-bands. Moreover, the required number of RIS elements can be reduced by increasing the BS transmit power.} %Besides, the number of RIS elements can also be reduced by deploying RIS closer to the BS, as when RIS is closer to the BS, the signal experiences a clear LoS channel which maximizes the received signal power, hence the target SE can be achieved with less RIS elements.

\section{Research Directions}
In this section, we present promising future research directions that we consider to be of great importance to unlock the full potential of RISs for 6G networks.

\subsection {RIS-assisted Terahertz (THz) Communication}
The THz communication, with ultra-wide bandwidth, is considered to be a promising candidate for 6G communication. Because of the ultra-high frequency, the THz signal may undergo severe signal attenuation and communication interruptions. To this end, RIS can be applied to THz communication for better coverage performance. However, the RIS-assisted THz communication poses a major challenge of exploiting the unique propagation properties of THz in the RIS-assisted network, which needs to be addressed.

\subsection {Aerial RIS Empowered Communication}
The aerial RIS, carried by UAV or balloon, can realize full-space reflections to serve a relatively larger number of users than the ground RIS, fixed at a location. The aerial RIS is more likely to enjoy the LoS channel conditions, thus mitigating the blockages. 
The high mobility of UAVs can be exploited to further expand the coverage of aerial RIS. However, in practice,  the aerial RIS brings new challenges, including the three-dimensional (3D) placement and the channel estimation, that are thus worthy of investigation.

\subsection {RIS-assisted Physical Layer Security}
The RIS's signal manipulation capability can even enhance the physical layer security (PLS) of the communication links, by simultaneously boosting the signal beam at the intended user and suppressing the beam at the unintended user. The RIS-assisted PLS requires the information of the channels from the eavesdropper to the RIS and the BS, which is difficult to obtain in practice. This, therefore, calls for a sophisticated channel estimation and RIS passive beamforming design for RIS-assisted PLS under imperfect CSI.

\subsection{\textcolor{black}{RIS-assisted Optical Wireless Communication}}
\textcolor{black}{Optical wireless communication (OWC) is a promising solution for next-generation high data rate applications at relatively low hardware cost and complexity than the RF counterpart. Nonetheless, the performance of OWC is subject to the existence of LoS between the transceivers. To relax this constraint, RIS can be applied to OWC to mitigate the LoS blockages by directing the optical beam in a desired direction. Thus, the integration of RIS and OWC can enable a plethora of applications for both indoor and outdoor scenarios.}

\subsection{\textcolor{black}{RIS-assisted mMIMO Network}}
\textcolor{black}{mMIMO, which is the extension of MIMO technology, is one of the  key enablers for dramatically improving the transmission gain and spectral efficiency. However, high hardware cost and power consumption are the fundamental limitations towards the practical implementation of mMIMO systems. Nevertheless, RIS can be integrated with mMIMO to provide required performance gains in an energy-efficient and cost-effective fashion. The low complexity algorithms for beamforming designs and resource allocation for RIS-assisted mMIMO systems need to be studied to achieve maximum performance. }

%\subsection{\textcolor{black}{Artificial Intelligence Driven techniques for RIS-assisted networks}}

\section{Conclusion}
%The RIS empowered wireless networks, with high energy and spectral efficiency, pave the way towards smart radio environment.
~\textcolor{black}{In this article, we overviewed the potentials of RIS technology for 6G wireless networks. We first discussed the performance gains that can be achieved by integrating RIS with emerging communication technologies, such as NOMA, SWIPT, UAVs, BackCom, mmWaves, and multi-antenna systems. Despite the great potentials, RIS encounters new challenges to be efficiently integrated into the wireless network. In this regard, we exposed the crucial challenges for the practical implementation of RIS-assisted networks. A case study for RIS-assisted NOMA network under imperfect CSI has also been presented to demonstrate the importance of better channel estimation for RIS-assisted networks and to indicate the various factors affecting the size of RIS. Finally, to provide effective guidance for future research, we highlighted promising research directions for RIS-assisted networks.}

%\bibliographystyle{IEEEtran}
%\bibliography{output.bbl}

\vspace{-2cm}
\begin{IEEEbiographynophoto}{Sarah Basharat}(sbasharat.msee19seecs@seecs.edu.pk) 
received her B.E. and M.S. degrees in electrical engineering from National University of Sciences and Technology (NUST), Pakistan, in 2019 and 2021, respectively. Her research interests include B5G and 6G communications, non-orthogonal multiple access (NOMA), and reconfigurable intelligent surfaces (RISs).
\end{IEEEbiographynophoto}
\vspace{-2cm}
\begin{IEEEbiographynophoto}{Syed Ali Hassan} [S’08, M’11, SM’17] (ali.hassan@seecs.edu.pk) received his Ph.D. in electrical engineering from Georgia Tech, Atlanta, in 2011, his M.S. in mathematics from Georgia Tech in 2011, and his M.S. in electrical engineering from the University of Stuttgart, Germany, in 2007, and a B.E. in electrical engineering (highest honors) from the National University of Sciences and Technology (NUST), Pakistan, in 2004. Currently, he is working as an associate professor at NUST, where he is heading the IPT research group, which focuses on various aspects of theoretical communications. 
\end{IEEEbiographynophoto}
\vspace{-2cm}
\begin{IEEEbiographynophoto}{Haris Pervaiz}[S’09, M’09] (h.b.pervaiz@lancaster.ac.uk) is currently an assistant professor with the School of Computing and Communications (SCC), Lancaster University, U.K. From April 2017 to October 2018, he was a research fellow with the 5G Innovation Centre, University of Surrey, U.K. From 2016 to 2017, he was an EPSRC Doctoral Prize Fellow with the SCC, Lancaster University. He received his Ph.D. degree from Lancaster University, U.K., in 2016. His current research interests include green heterogeneous wireless communications and networking, 5G and beyond, millimeter wave communication, and energy and spectral efficiency.
\end{IEEEbiographynophoto}
\vspace{-2cm}
\begin{IEEEbiographynophoto}{Aamir Mahmood}[M’18, SM’19] (aamir.Mahmood@miun.se) is an assistant professor of communication engineering at Mid Sweden University, Sweden. He received the M.Sc. and D.Sc. degrees in communications engineering from Aalto University School of Electrical Engineering, Finland, in 2008 and 2014, respectively. He was a research intern at Nokia Researcher Center, Finland and a visiting researcher at Aalto University during 2014-2016. His research interests include network time synchronization, resource allocation for URLLC, and RF interference/coexistence management.
\end{IEEEbiographynophoto}
\vspace{-2cm}
\begin{IEEEbiographynophoto}{Zhiguo Ding}[S’03, M’05, SM’17, F’20] (zhiguo.ding@manchester.ac.uk) is currently a Professor at the University of Manchester. Dr Ding research interests are 5G networks, signal processing and statistical signal processing. He has been serving as an Editor for IEEE TCOM, IEEE TVT, and served as an editor for IEEE WCL and IEEE CL. He received the EU Marie Curie Fellowship 2012-2014, IEEE TVT Top Editor 2017, 2018 IEEE COMSOC Heinrich Hertz Award, 2018 IEEE VTS Jack Neubauer Memorial Award, and 2018 IEEE SPS Best Signal Processing Letter Award.
\end{IEEEbiographynophoto}
\vspace{-2cm}
\begin{IEEEbiographynophoto}{Mikael Gidlund}[M’98, SM’16] (mikael.gidlund@miun.se) is a professor of computer engineering at Mid Sweden University, Sweden. He has worked as Senior Principal Scientist and Global Research Area Coordinator of Wireless Technologies, ABB Corporate Research, Sweden, Project Manager and Senior Specialist with Nera Networks AS, Norway, and Research Engineer and Project Manager with Acreo AB, Sweden. His current research interests include wireless communication and networks, wireless sensor networks, access protocols, and security. 
\end{IEEEbiographynophoto}

\end{document}